\documentclass[nofootinbib,superscriptaddress]{revtex4}

\begin{document}

  \title{Quantum Physics from A to Z}

  \author{M. Arndt}
  \affiliation{University of Vienna, Vienna, Austria}

  \author{M. Aspelmeyer}
  \affiliation{University of Vienna, Vienna, Austria}

  \author{H. J. Bernstein}
  \affiliation{Hampshire College, Amherst, MA, USA}

  \author{R. Bertlmann}
  \affiliation{University of Vienna, Vienna, Austria}

  \author{C. Brukner}
  \affiliation{University of Vienna, Vienna, Austria}

  \author{J. P. Dowling}
  \affiliation{Louisiana State University, Baton Rouge, LA, USA}

  \author{J. Eisert}
  \affiliation{University of Potsdam, Potsdam, Germany}

  \author{A. Ekert}
  \affiliation{Cambridge University, Cambridge, UK}

  \author{C. A. Fuchs}
  \affiliation{Lucent Technologies, Murray Hill, NJ, USA}

  \author{D. M. Greenberger}
  \affiliation{City College, CUNY, NY, USA}

  \author{M. A. Horne}
  \affiliation{Stonehill College, Easton, MA, USA}

  \author{T. Jennewein}
  \affiliation{Austrian Academy of Sciences, Vienna, Austria}

  \author{P. G. Kwiat}
  \affiliation{University of Illinois, Urbana-Champaign, IL, USA}

  \author{N. D. Mermin}
  \affiliation{Cornell University, Ithaca, NY, USA}

  \author{J.-W. Pan}
  \affiliation{University of Heidelberg, Heidelberg, Germany}

  \author{E. M. Rasel}
  \affiliation{University of Hannover, Hannover, Germany}

  \author{H. Rauch}
  \affiliation{Atomic Institute of the Austrian Universities, Vienna,
  Austria}

  \author{T. G. Rudolph}
  \affiliation{Imperial College, London, United Kingdom}

  \author{C. Salomon}
  \affiliation{Ecole Normale Superieure, Paris, France}

  \author{A. V. Sergienko}
  \affiliation{Boston University, Boston, MA, USA}

  \author{J. Schmiedmayer}
  \affiliation{University of Heidelberg, Heidelberg, Germany}

  \author{C. Simon}
  \affiliation{CNRS, Grenoble, France}

  \author{V. Vedral}
  \affiliation{University of Leeds, Leeds, UK}

  \author{P. Walther}
  \affiliation{University of Vienna, Vienna, Austria}

  \author{G. Weihs}
  \affiliation{University of Waterloo, Waterloo, Canada}

  \author{P. Zoller}
  \affiliation{University of Innsbruck, Innsbruck, Austria}

  \author{M. Zukowski}
  \affiliation{University of Gdansk, Gdansk, Poland}

  \date{May 20$^{th}$, 2005}

  \begin{abstract}
  This is a collection of statements gathered on the occasion of
  the \textit{Quantum Physics of Nature} meeting in Vienna.
    \end{abstract}

  \maketitle

Since philosophers are beginning to discuss the ``Zeilinger
principle'' of quantum physics \cite{nature,dpg,wikipedia} it
appears timely to get a community view on what that might be. The
interpretation in question seems to have been developed over a
period totalling 60 years to the day and does not only concern
quantum mechanics but physics as a whole.

We begin to shed some light on this issue with a grossly incomplete
collection of statements attributed to but not necessarily
authorized by Anton Zeilinger.

\begin{itemize}
  \item There is no quantum information. There is only a quantum way
  of handling information.

  \item Photons are clicks in photon detectors.

  \item Descriptions of experiments with photons should not make use
  of the term vacuum.

  \item Objective randomness is probably the most important element
  of quantum physics.

  \item One elementary quantum system can carry one bit of
  information.

  \item The world is quantized because information is quantized.

  \item Quantum mechanics applies to single quantum systems.

  \item The border between classical and quantum phenomena is just a
  question of money.


  \item All the easy experiments have been done.

  \item Good papers have to be sexy.

\end{itemize}

In order to stimulate a discussion we now present our personal views
relating to Anton Zeilinger's philosophy of physics and some
personal recollections in alphabetic author order.

\textbf{Markus Arndt}:  There will be an object size where the
monetary value of all countries on earth will not suffice to isolate
the quantum system well enough. An upper limit is certainly earth
itself. The suppression of decoherence for viruses or bacteria is
probably indeed a question of money and technological innovations
alone. The quantum nature of  objective randomness is mirrored by
the fact that it troubles and pleases me at the same time. I learned
from A.Z. to appreciate the use of operational and verifiable
definitions of terms. Applied with consequence it may however lead
to contradictions with other statements we like to accept. Objective
(absolute) randomness is hardly fully verifiable. One may exclude
certain classes of causes / reasons for a quantum choice.
Statistical tests on results of quantum measurements appear to be
perfect. But the a priori exclusion of any reason whatsoever cannot
be falsified / verified. The German language allows to make the
distinction between ``Realit\"at'' and ``Wirklichkeit''. Realit\"at
relates to the Latin word for thing (res). Wirklichkeit contains the
notion of effecting something (``Wirkung''). I prefer
``Wirklichkeit'' because I think that no object can be defined
without reference to its external world, which is in perpetual
change.

\textbf{Markus Aspelmeyer}: Quantum physics describes physical
situations, in which information is fundamentally bounded. Future
experiments will show that this principle will carry through also
for macroscopic systems, if they are properly prepared. The
classical-quantum boundary is simply a matter of information
control. The fascination of quantum physics lies in the fact that it
is not yet formulated coherently by a set of simple foundational
principles. I do believe that A.Z.'s foundational principle is the
most beautiful approach that we currently have; I am also convinced
that the next step will require a bold "creative act" to establish a
way to talk quantum physics in terms of information - similar to
Newton's introduction of "momentum" and "force" (at his time terms
from the colloquial language) to talk classical physics.

\textbf{Herbert Bernstein}: A.Z.'s suggestion that quantum
information is a misnomer intrigues me. I see  three contending
metaphors -- each with leading proponents. These are ``the creation
of meaning'', ``the thermodynamic analogy built on quantum entropy
measures'', and the ``resource (commodity) model.'' Perhaps A.Z.'s
remark is closest to the first of these. I look forward to seeing
which becomes paradigmatic as the millennium unfolds.

\textbf{Reinhold Bertlmann}: 1) The actual reality lies beyond the
socks! 2) I remember John Bell saying: ``There is no quantum
logic''. 3) The true quantum mystery starts with more than one
particle. 4) There is no border between classical and quantum
phenomena -- you just have to look closer. 5) Photons are vibrations
of the vacuum. 6) Yes, papers should be sexy -- but the authors too!
(Greenberger-Bertlmann theorem)

\textbf{Caslav Brukner}: Everything that physics is about is
extracting meaning whatsoever out of data of our experience. But the
ultimate experience is nothing more than a stream of "yes" and/or
"no" answers to the questions posed to Nature. I do agree with AZ:
What could be more natural than to assume that the most elementary
system from which we intend to extract meaning can give answer to
only one single question?

\textbf{John Dowling}: There are no sexy papers -- there is only a
sexy way of handling homely papers!

\textbf{Jens Eisert}: 1) That Einstein's programme put forth in his
EPR paper has finally failed does not mean that he did not point
towards the aspect of quantum mechanics where it fundamentally
departs from classical physics. 2) Fundamental question about de
Broglie wave interferometry: What's the matter, dude?

\textbf{Artur Ekert}: 1) Good titles get long mileage. 2) I like
small gadgets, look at this tiny digital camera... where is it...

\textbf{Chris Fuchs}: I used the following quote to open my talk. It
comes from Anton taking part in in a panel discussion at the
Symposium on the Foundations of Modern Physics 1994 in Helsinki:

``We don't know why events happen, as expressed by Bell. Let me
explain a little bit what I mean by that.  By quantum phenomenon we
mean the whole unity from preparation via evolution and propagation
to detection.  Then there is an uncontrollable element somewhere in
this chain.  It can be called the reduction of the wave packet.  Or
it can be in the many worlds interpretation the unexplainability of
the fact that I find my consciousness in one given universe and not
in the others.  In a Bohm interpretation it can be the fact that I
cannot control the initial condition.  As an experimentalist I would
say that there is some uncontrollable element from the following
point of view. When doing a Stern-Gerlach experiment, for example,
with an x-polarized spin, I cannot predict that this spin will go
up, this one will go down, etc.  There is something beyond my
control. My personal feeling is that we have found for the first
time in physics that there are things which happen without
sufficient reason.  This, I think, is a very profound discovery.  I
don't know whether there is a way to understand this or not.  I feel
there might be a way to understand why the world is so strange but
we have not understood that yet.  In my opinion this so, because we
really don't know what information is.  We don't know what it means
to collect information about the world.  There is some world out
there.  In the words of Professor Laurikainen, in a very specific
sense we have created the whole universe.  But in some sense it
exists without us. We have to understand therefore what it means to
collect information about something which is not as much structured
as we think.''

\textbf{Dan Greenberger}: I must disagree with A.Z. in that I
believe that there is \textbf{no} classical world. There is
\textbf{only} a quantum world. Classical physics is a collection of
unrelated insights: Newton´s laws, Hamilton´s principle, etc. Only
quantum theory brings out their connection. An analogy is the
Hawaiian Islands, which look like a bunch of islands in the ocean.
But if you could lower the water, you would see, that they are the
peaks of a chain of mountains. That is what quantum physics does to
classical physics. " Quantum mechanics is magic! It is not black
magic, but it is nonetheless magic!

\textbf{Mike Horne}: Who needs creation operators; we've got
down-conversion. Who needs creation operators; we've got the
amplitudes! Physics is phun! If it's simple, it probably has not
been done!

\textbf{Thomas Jennewein}: Quoting A.Z.: \emph{1) Photons are just
clicks in photon detectors; nothing real is traveling from the
source to the detector. 2) Their coherence length is only a concept
that describes their ability to interfere, nothing more, like the
common belief of a ``region of space with a fixed phase relation''.
2) Maybe it is not ``bit from it'' but ``it from bit''?} --- But
what about the energy flowing from the source to the detector?

\textbf{Paul Kwiat}: A.Z.'s first advice to me: \textit{The best way
to learn a new language is to sleep with a dictionary.} A.Z.'s best
advice to me, after I commented a perpetual feeling of being behind:
\textit{You will never catch up. So don't bother trying. Have fun!}
My own advice: The words ``sinnvoll'' and ``sinnlich'' are
\emph{not} interchangeable.

\textbf{N. David Mermin}: Last year I had a column in Physics Today
on whether Bohr actually said what he is said to have said, and
another column on how Feynman was unlikely to have said what he is
said to have said (because, actually, I said it). So I view with
great suspicion a list of what Anton Zeilinger is said to have said.
With that clearly understood, here are some comments:

``There is no quantum information. There is only a quantum way of
treating classical information.'' I entirely agree. This is very
close to the point of my talk at the QUPON conference. I would only
add that, at least for quantum computers, that classical information
is entirely restricted to the readings of initial, intermediate, and
final one-qubit measurement gates (and the structure of the quantum
circuit).

``Photons are clicks in photon counters.'' A special case of Aage
Bohr and Ole UIfbeck's rule that there are no particles, only
clicks. Would Anton agree that electrons are clicks in electron
counters? Are fullerenes clicks in fullerene counters? Is Anton a
click in an Anton counter?

``Descriptions of experiments with photons should not make use of
the term vacuum.'' Thank you, Anton.  When I get to the term
``vacuum'' I stop reading because I know the paper is about to
became unintelligible, and that if l have to understand it, I'll be
able to find somebody who can explain it  in clearer terms.

``All the easy experiments have been done.'' I wouldn't know. But
beware of hard experiments whose only interest is that they are
hard. It's the experimentalist's version of finding an exact
solution to a model whose only interest is that it has an exact
solution.

``Papers should be sexy.'' I would have said ``entertaining'' on the
ground that sex is not the only way to have fun. Unreadable or just
dull papers are the plague of our time.

``Objective randomness is probably the most important element of
quantum physics'' Why ``probably''? Is this uncharacteristic
cautiousness, or just a B+ joke? I used to say the same thing, but
Chris Fuchs has taught me to beware of conjoining ``objective'' to
``probability''. These days I'd rather say that a great lesson of
quantum physics is that our knowledge can only be probabilistic.

\textbf{Jian-Wei Pan}: Confucius says, good is the advisor who leads
young experimentalists, but better is the one who can transform a
theorist into an experimentalist.

\textbf{Ernst Rasel}: Information is everything, also in the case of
cooking. (From a cookbook given to the author on the occasion of his
graduation by A.Z.)

\textbf{Helmut Rauch}: 1) Re randomness: Schr\"odinger's equation is
deterministic, but we don't know exactly the interaction Hamiltonian
and boundary conditions. 2) Re photons: When the photons travel
through the universe they are existent as well. 3) Reality is much
more than information, which is much more than knowledge, which is
much more than understanding, etc.

\textbf{Terry Rudolph}: Without Zeilingers, Vedrals and other
glorified chimpanzees there is no need for information whatsoever.

\textbf{Christophe Salomon}: I believe there is a lot to discover at
the frontier of quantum physics and gravitation. After all, we
understand very little about our universe!

\textbf{J\"org Schmiedmayer}: It does not help to close your eye in
front of quantum systems. If we don't know the outcome of an
experiment it does not matter, if the demon can know it, that's
enough. Quantum physics reflects exactly the way we can see nature
in our (optimized) experiments.  It reflects exactly the maximum
information we {\em can possibly} get out from nature.  If we want
to look beyond, we have to find a completely new way to look at
nature.

Quoting A.Z. on unscientific business: \textit{I don't know, I just
work here}.

\textbf{Alexander Sergienko}: Only detected photons are good
photons. There are many theoretical models describing a physical
system but only one experimental observation.

\textbf{Christoph Simon}: There are many orders of magnitude between
the microscopic and macroscopic that are unexplored from the point
of view of fundamental quantum effects such as superpositions and
entanglement. (Compare a C$_{60}$ molecule, which consists of
$\approx 10^3$ nucleons and a cat, which has about $10^{27}$ of
them.) It is a fascinating challenge to try to push the borders of
the region within which quantum effects have been observed. Is the
observation of quantum phenomena in larger and larger systems really
just a ``question of money''? Or are there fundamental limitations
to the validity of quantum physics as we know it? There should be
enough work for a few generations of physicists in trying to answer
these questions.

\textbf{Vlatko Vedral}: There is no classical information. The
entropy of the universe is constant (zero)\footnote{$\infty$ (V.
Buzek)},\footnote{42 (T. Rudolph)}. There is no quantum measurement.
Reversibility is just a question of money

\textbf{Philip Walther}: Quantum mechanics seems to be nature's
organized self-defense for not revealing its inner secrets. Why are
coworkers never sexy?

\textbf{Gregor Weihs}: If the two-state system is the elementary
quantum system as postulated by the informational principle, can a
qutrit be elementary at all? A qutrit cannot be decomposed trivially
into qubits. Very often qutrits appear as the symmetric subspace of
two qubits, but then there must exist a singlet as well, possibly
with very different properties. The only elementary three-state
systems we know are the vector bosons of the strong and weak
interactions (the photon being counted as two-state here). I wonder
whether the vector bosons might be composite as well.

\textbf{Peter Zoller}: Quoting A.Z.: \emph{Physics will in the
future put less emphasis on equations and mathematics but more on
verbal understanding.}

\textbf{Marek Zukowski}: The title of our paper is sexy, but not
sexy enough!


\begin{thebibliography}{10}
  \bibitem{nature} Q. Schiermeier, \textit{Quantum physics:  The philosopher of photons,} Nature 434, 1066-1066 (2005).
  \bibitem{dpg} H. Hille, \textit{Zeilinger und die Entdeckung des Subjekts. Sein Buch Einsteins
  Schleier}, Abstract AKPHIL 5.1, German Physical Society Spring Meeting, Berlin 2005.
  \bibitem{wikipedia} http://en.wikipedia.org/wiki/Information\_Theory
\end{thebibliography}
\end{document}